\documentclass[preprint,showpacs,preprintnumbers,amsmath,amssymb,superscriptaddress]{revtex4}

\usepackage{times}
\usepackage{graphicx}% Include figure files
\usepackage{dcolumn}% Align table columns on decimal point
\usepackage{bm}% bold math
\usepackage{natbib}
\begin{document}

\title{Observation of ferromagnetic resonance in strontium ruthenate (SrRuO$_{3}$)}% Force line breaks with \\

\author{M.C. Langner}
\affiliation{Department of Physics, University of California, Berkeley, CA 94720}
\affiliation{Materials Science Division, Lawrence Berkeley National Laboratory, Berkeley, CA 94720}
\author{C.L.S. Kantner}
\affiliation{Department of Physics, University of California, Berkeley, CA 94720}
\affiliation{Materials Science Division, Lawrence Berkeley National Laboratory, Berkeley, CA 94720}
\author{Y.H. Chu}
\affiliation{Department of Materials Science and Engineering, National Chiao Tung University, HsinChu, Taiwan, 30010}
\author{L.M. Martin}
\affiliation{Materials Science Division, Lawrence Berkeley National Laboratory, Berkeley, CA 94720}
\author{P. Yu}
\affiliation{Department of Physics, University of California, Berkeley, CA 94720}
\author{R. Ramesh}
\affiliation{Department of Physics, University of California, Berkeley, CA 94720}
\affiliation{Department of Materials Science and Engineering, University of California, Berkeley, CA 94720}
\author{J. Orenstein}
\affiliation{Department of Physics, University of California, Berkeley, CA 94720}
\affiliation{Materials Science Division, Lawrence Berkeley National Laboratory, Berkeley, CA 94720}

\date{\today}% It is always \today, today,

\begin{abstract}
We report the observation of ferromagnetic resonance (FMR) in SrRuO$_{3}$ using the time-resolved magneto-optical Kerr effect. The FMR oscillations in the time-domain appear in response to a sudden, optically induced change in the direction of easy-axis anistropy.  The high FMR frequency, 250 GHz, and large Gilbert damping parameter, $\alpha\approx$ 1, are consistent with strong spin-orbit coupling.  We find that the parameters associated with the magnetization dynamics, including $\alpha$, have a non-monotonic temperature dependence, suggestive of a link to the anomalous Hall effect.\end{abstract}

\maketitle

Understanding and eventually manipulating the electron's spin degree of freedom is a major goal of contemporary condensed matter physics.  As a means to this end, considerable attention is focused on the spin-orbit (SO) interaction, which provides a mechanism for control of spin polarization by applied currents or electric fields \cite{spintronics}. Despite this attention, many aspects of SO coupling are not fully understood, particularly in itinerant ferromagnets where the same electrons are linked to  both rapid current fluctuations and slow spin dynamics. In these materials, SO coupling is responsible for spin-wave damping \cite{kambersky, korenman}, spin-current torque \cite{slonc, berger}, the anomalous Hall effect (AHE) \cite{karplus}, and magnetocrystalline anisotropy (MCA) \cite{MCA}. Ongoing research is aimed toward a quantitative understanding of how bandstructure, disorder, and electron-electron interactions interact to determine the size and temperature dependence of these SO-driven effects.

SrRuO$_3$ (SRO) is a material well known for its dual role as a highly correlated metal and an itinerant ferromagnet with properties that reflect strong SO interaction \cite{anisotropydir, irconductivity, marshall}.  Despite its importance as a model SO-coupled system, there are no previous reports of ferromagnetic resonance (FMR) in SRO. FMR is a powerful probe of SO coupling in ferromagnets, providing a means to measure both MCA and the damping of spin waves in the small wavevector regime \cite{heinrich}. Here we describe detection of FMR by time-resolved magnetooptic measurements performed on high-quality SRO thin films.  We observe a well-defined resonance at a frequency $\Omega_{FMR}$ = 250 GHz.  This resonant frequency is an order of magnitude higher than in the transition metal ferromagnets, which accounts for the nonobservation by conventional microwave techniques.

10-200 nm thick SRO thin films were grown via pulsed laser deposition between 680-700$^{\circ}$C in 100 mTorr oxygen. High-pressure reflection high-energy electron diffraction (RHEED) was used to monitor the growth of the SRO film in-situ.  By monitoring RHEED oscillations, SRO growth was determined to proceed initially in a layer-by-layer mode before transitioning to a step-flow mode. RHEED patterns and atomic force microscopy imaging confirmed the presence of pristine surfaces consisting of atomically flat terraces separated by a single unit cell step (~3.93 {\AA}). X-ray diffraction indicated fully epitaxial films and x-ray reflectometry was used to verify film thickness.  Bulk magnetization measurements using a SQUID magnetometer indicated a Curie temperature, T$_{C}$, of $\sim$ 150K.
       
Sensitive detection of FMR by the time-resolved magnetooptic Kerr effect (TRMOKE) has been demonstrated previously \cite{precession1, precession2, precession0}.  TRMOKE is an all optical pump-probe technique in which the absorption of an ultrashort laser pulse perturbs the magnetization, $\textbf{M}$, of a ferromagnet.  The subsequent time-evolution of $\textbf{M}$ is determined from the polarization state of a normally incident, time-delayed probe beam that is reflected from the photoexcited region. The rotation angle of the probe polarization caused by absorption of the pump, $\Delta \Theta_{K}(t)$, is proportional to $\Delta M_{z}(t)$, where $z$ is the direction perpendicular to the plane of the film \cite{magnetooptics}.

Figs. 1a and 1b show $\Delta \Theta_{K}(t)$ obtained on an SRO film of thickness 200 nm.  Very similar results are obtained in films with thickness down to 10 nm.  Two distinct types of dynamics are observed, depending on the temperature regime.  The curves in Fig. 1a were measured at temperatures near T$_{C}$. The relatively slow dynamics agree with previous reports for this T regime \cite{ogasawara} and are consistent with critical slowing down in the neighborhood of the transition \cite{slowingdown}. The amplitude of the photoinduced change in magnetization has a local maximum near T=115 K.  Distinctly different magnetization dynamics are observed as T is reduced below about 80 K, as shown in Fig. 1b.  The TRMOKE signal increases again, and damped oscillations with a period of about 4 ps become clearly resolved.

\begin{figure}[h]
\label{fig:fig1}
\includegraphics[width=2.75in]{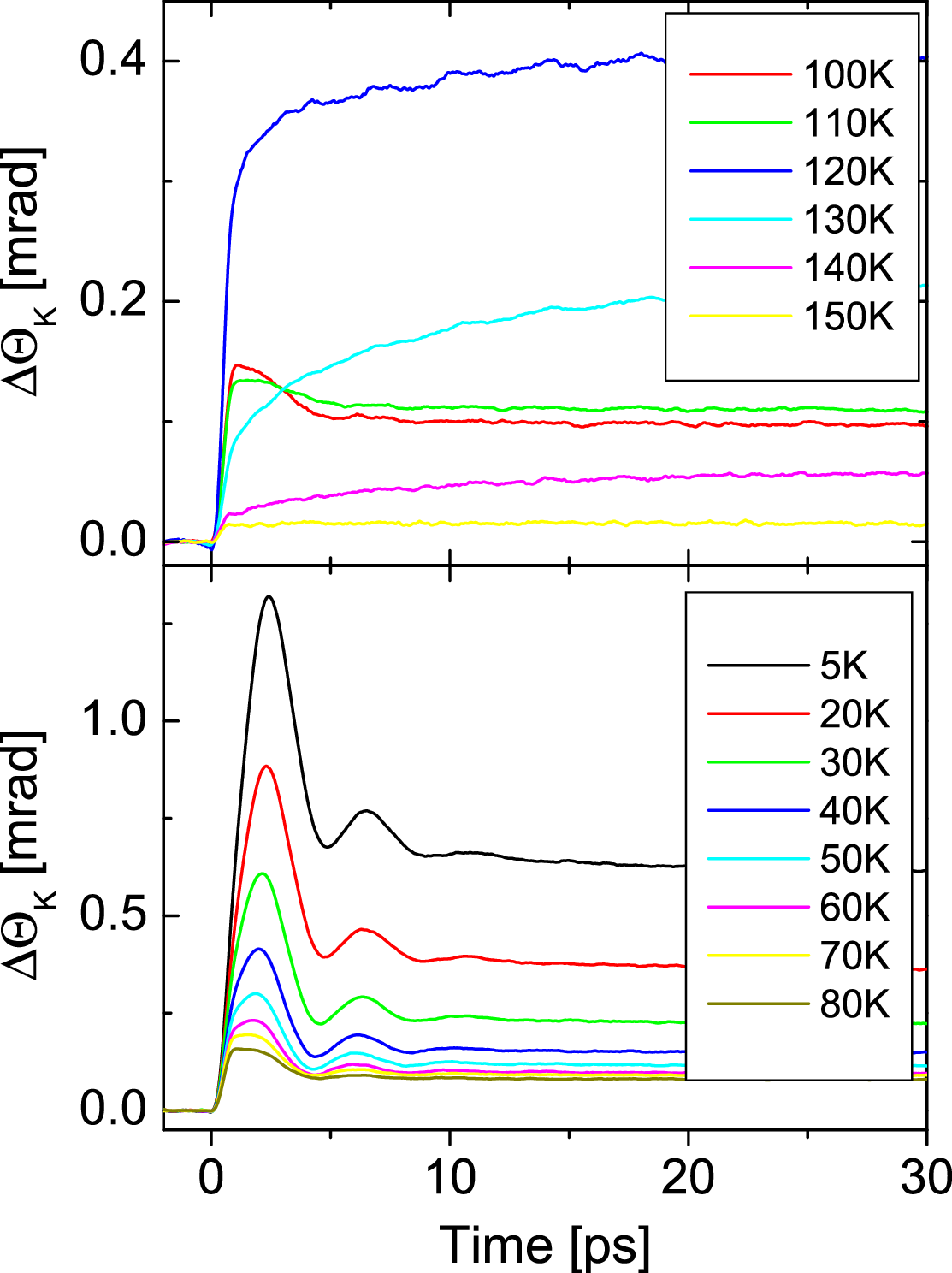}
\caption{Change in Kerr rotation as a function of time delay following pulsed photoexcitation, for several temperatures below the Curie temperature of 150 K. Top Panel: Temperature range 100 K $<$T $<$ 150 K.   Bottom panel:  Temperature range 5 K $<$T $<$ 80 K. Signal amplitude and oscillations grow with decreasing T.}
\end{figure}

In order to test if these oscillations are in fact the signature of FMR, as opposed to another photoinduced periodic phenomenon such as strain waves, we measured the effect of magnetic field on the TRMOKE signals.  Fig. 2a shows $\Delta \Theta_{K}(t)$ for several fields up to 6 T applied normal to the film plane.  The frequency clearly increases with increasing magnetic field, confirming that the oscillations are associated with FMR.

The mechanism for the appearance of FMR in TRMOKE experiments is well understood \cite{precession0}.  Before photoexcitation, \textbf{M} is oriented parallel to \textbf{h$_{A}$}.  Perturbation of the system by the pump pulse (by local heating for example) generates a sudden change in the direction of the easy axis.  In the resulting nonequilibrium state, \textbf{M} and \textbf{h$_{A}$} are no longer parallel, generating a torque that induces \textbf{M} to precess at the FMR frequency.  In the presence of Gilbert damping, \textbf{M} spirals towards the new \textbf{h$_{A}$}, resulting in the damped oscillations of $M_z$ that appear in the TRMOKE signal.

To analyze the FMR further we Fourier transform (FT) the time-domain data and attempt to extract the real and imaginary parts of the transverse susceptibility, $\chi_{ij} \left( \omega \right)$. The magnetization in the time-domain is given by the relation,
\begin{equation}
\Delta M_{i}\left(t\right) = \int^{\infty}_{0} \chi_{ij} \left( \tau \right) \Delta h^{j}_{A} \left(t-\tau \right) d\tau,
\end{equation}
where $\chi_{ij}(\tau)$ is the impulse response function and $\Delta \boldsymbol{h_{A}}(t)$ is the change in anisotropy field. If $\Delta \boldsymbol{h_{A}}(t)$ is proportional to the $\delta$-function, $\Delta M_{i}(t)$ is proportional $\chi_{ij}(\tau)$ and the FT of the TRMOKE signal yields $\chi_{ij}(\omega)$ directly. However, for laser-induced precession one expects that $\Delta \boldsymbol{h_{A}}(t)$ will be more like the step function than the impulse function, as photoinduced local heating can be quite rapid compared with cooling via thermal conduction from the laser-excited region.  When $\Delta \boldsymbol{h_{A}}(t)$ is proportional to the step function, $\chi_{ij} (\omega)$ is proportional to the FT of the time derivative of $\Delta M_{i}(t)$, rather than $\Delta M_{i}(t)$ itself.  In this case, the observable $\omega \text{Im} \left\{ \Delta \Theta_{K} ( \omega ) \right\}$ should be closely related to the real, or dissipative part of $\chi_{ij}(\omega)$.

\begin{figure}
\label{fig:fig2}
\includegraphics[width=2.75in]{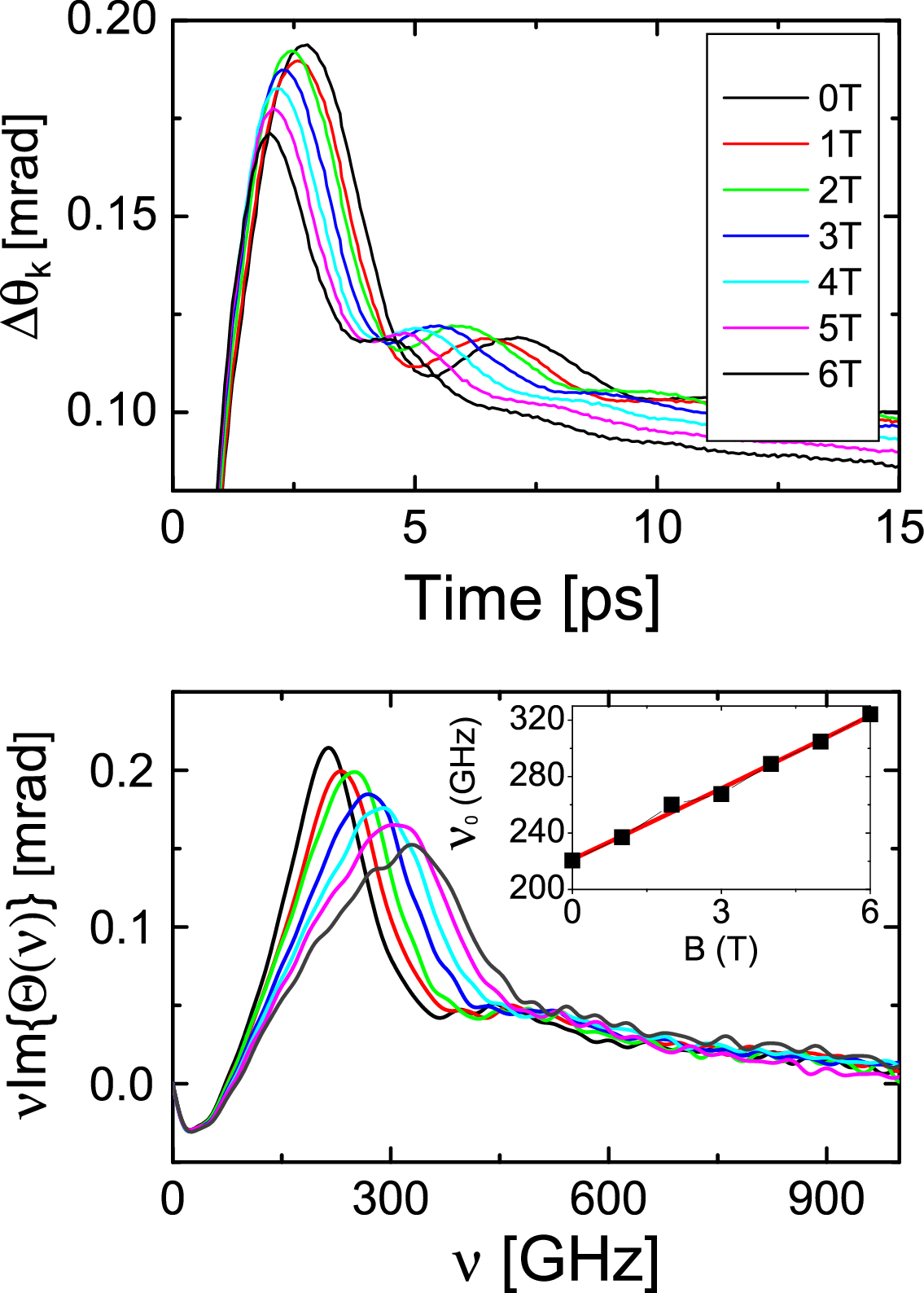}
\caption{Top panel: Change in Kerr rotation as a function of time delay following pulsed photoexcitation at T=5 K, for several values of applied magnetic field ranging up to 6 Tesla. Bottom panel: Fourier transforms of signals shown in top panel.  Inset: FMR frequency vs. applied field.}
\end{figure}

In Fig. 2b we plot $\omega \text{Im} \left\{ \Delta \Theta_{K} ( \omega ) \right\}$ for each of the curves shown in Fig. 2a.  The spectra shown in Fig. 2b do indeed exhibit features that are expected for Re $\chi_{ij} ( \omega )$  near the FMR frequency.  A well-defined resonance peak is evident, whose frequency increases with magnetic field as expected for FMR.  The inset to Fig. 2b shows $\Omega_{FMR}$ as a function of applied magnetic field. The solid line through the data points is a fit obtained with parameters $\left| \boldsymbol{h_{A}} \right| = $ 7.2 T and easy axis direction equal to 22 degrees from the film normal.  These parameter values agree well with previous estimates based on equilibrium magnetization measurements \cite{anisotropydir, marshall}.

\begin{figure}
\label{fig:fig3}
\includegraphics[width=2.75in]{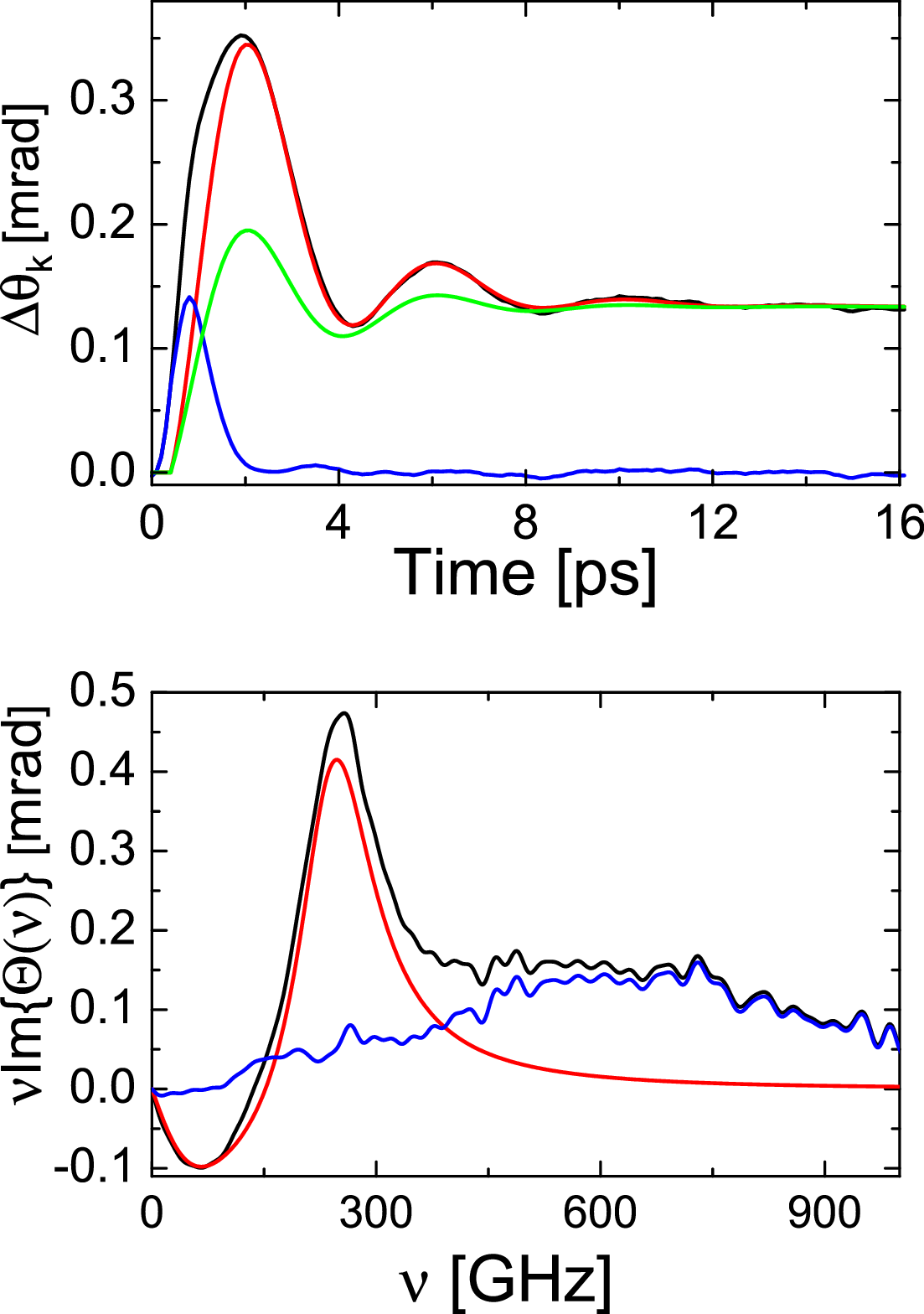}
\caption{Components of TRMOKE response in time (top panel) and frequency (bottom panel) domain. Black lines are the observed signals. Green line in the top panel is the simulated response to a step function change in easy-axis direction. Best fits to the "overshoot" model described in the text are shown in red. Blue lines are the difference between the measured and best-fit response.}
\end{figure}

Although the spectra in Fig. 2b are clearly associated with FMR, the sign change at low frequency is not consistent with Re $\chi_{ij} ( \omega )$, which is positive definite.  We have verified that the negative component is always present in the spectra and is not associated with errors in assigning the $t$=0 point in the time-domain data.  The origin of negative component of the FT is made clearer by referring back to the time domain.  In Fig. 3a we show typical time-series data measured in zero field at 5 K. For comparison we show the response to a step function change in the easy axis direction predicted by the Landau-Lifschitz-Gilbert (LLG) equations \cite{LLG}.  It is clear that, if the measured and simulated responses are constrained to be equal at large delay times, the observed $\Delta \Theta_{K} (t)$ is much larger than the LLG prediction at small delay.

We have found that $\Delta \Theta_{K}(t)$ can be readily fit by LLG dynamics if we relax the assumption that $\Delta \boldsymbol{h_{A}} (t)$ is a step-function, in particular by allowing the change in easy axis direction to "overshoot" at short times.  The overshoot suggests that the easy-axis direction changes rapidly as the photoexcited electrons approach quasiequilibrium with the phonon and magnon degrees of freedom.  The red line in Fig. 3a shows the best fit obtained by modeling $\Delta \boldsymbol{h_{A}} (t)$ by $H(t)(\phi_{0} + \phi_{1} e^{-t / \tau}$), where $H(t)$ is the step function, $\phi_0+\phi_1$ is the change in easy-axis direction at $t=0$, and $\tau$ is the time constant determining the rate of approach to the asymptotic value $\phi_0$.  The fit is clearly much better when the possibility of overshoot dynamics in $\Delta \boldsymbol{h_{A}}(t)$ is included.  The blue line shows the difference between measured and simulated response.  With the exception of this very short pulse centered near $t$=0, the observed response is now well described by LLG dynamics.  In principle, an alternate explanation for the discrepancy with the step-function assumption would be to consider possible changes in the magnitude as well as direction of \textbf{M}. However, we have found that fitting the data then requires $\left| \boldsymbol{M} ( t ) \right|$ to be larger at $t > 20$ ps than $\left| \boldsymbol{M} ( t < 0 ) \right|$, a photoinduced increase that is unphysical for a system in a stable FM phase.

In Fig. 3b we compare data and simulated response in the frequency domain.  With the allowance for an overshoot in $\Delta \boldsymbol{h_{A}} ( t )$  the spectrum clearly resolves into two components.  The peak at 250 GHz and the sign change at low frequency are the both part of the LLG response to $\Delta \boldsymbol{h_{A}} ( t )$.  The broad peak or shoulder centered near 600 GHz is the FT of the short pulse component shown in Fig. 3a.  We have found this component is essentially linear in pump pulse intensity, and independent of magnetic field and temperature - observations that clearly distinguish it from the FMR response.  Its properties are consistent with a photoinduced change in reflectivity due to band-filling, which is well-known to cross-couple into the TRMOKE signal of ferromagnets \cite{koopmans}.

By including overshoot dynamics in $\Delta \boldsymbol{h_{A}}(t)$, we are able to distinguish stimulus from response in the observed TRMOKE signals.  Assuming LLG dynamics, we can extract the two parameters that describe the response: $\Omega_{FMR}$ and $\alpha$; and the two parameters that describe the stimulus: $\phi_{1} / \phi_{0}$ and $\tau$.   In Fig. 4 we plot all four parameters as a function of temperature from 5 to 80 K. The T-dependence of the FMR frequency is very weak, with $\Omega_{FMR}$ deviating from 250 GHz by only about 5 $\%$ over the range of the measurement.  The Gilbert damping parameter $\alpha$ is of order unity at all temperatures, a value that is approximately a factor $10^{2}$ larger than found in transition metal ferromagnets.  Over the same T range the decay of the easy axis overshoot varies from about 2 to 4 ps. We note that the dynamical processes that characterize the response all occur in strongly overlapping time scales, that is the period and damping time of the FMR, and the decay time of the \textbf{h$_{A}$} overshoot, are each in the 2-5 ps range.

\begin{figure}[h]
\includegraphics[width=3.25in]{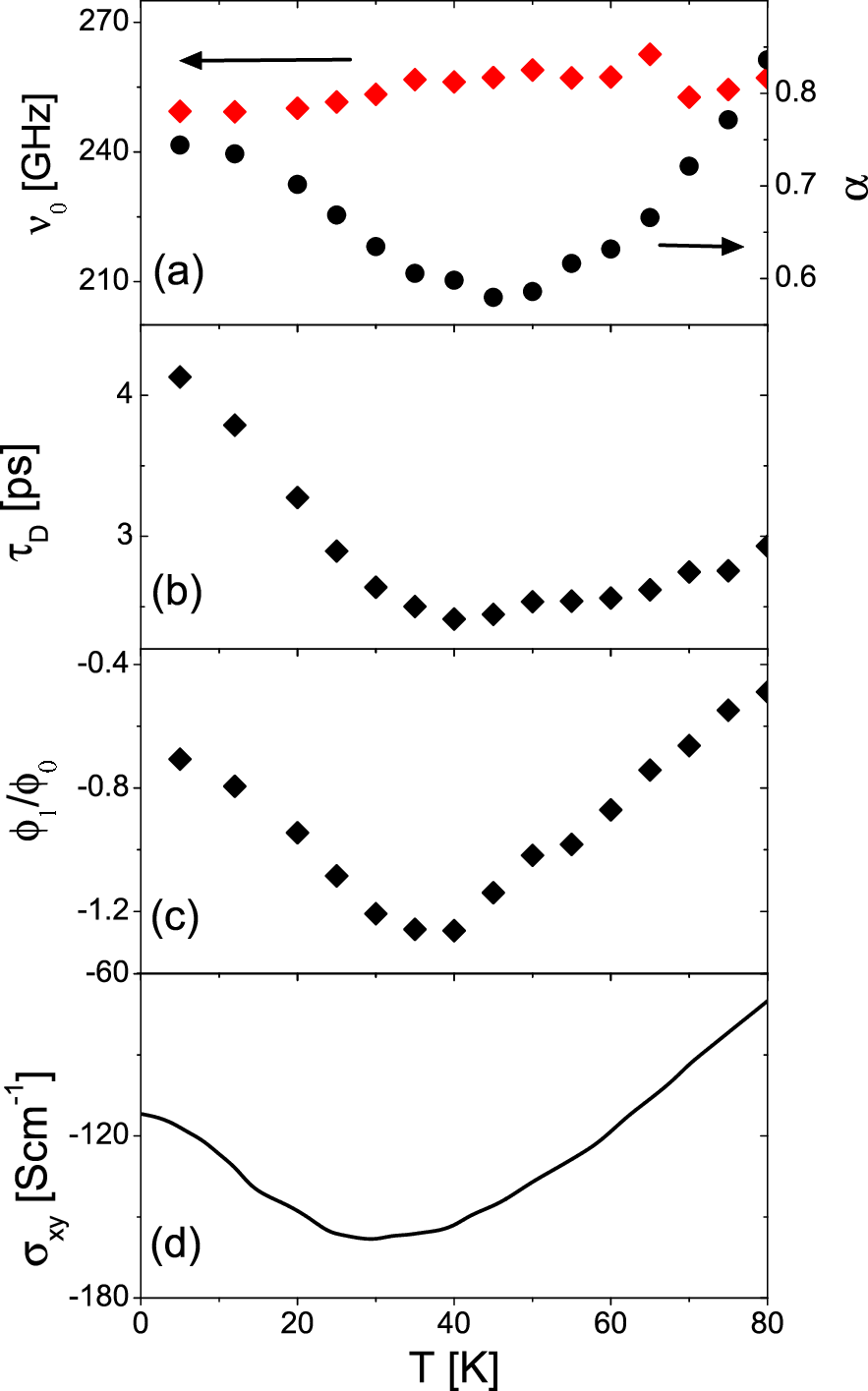}
\caption{Temperature dependence of (a) FMR frequency (triangles) and damping parameter (circles), (b) overshoot decay time, (c) ratio of overshoot amplitude to step-response amplitude ($\phi_{1} / \phi_{0}$), and (d) $\sigma_{xy}$ (adapted from \cite{mathieu}).}
\end{figure}

While $\Omega_{FMR}$ is essentially independent of T, the parameters $\alpha$, $\phi_{1} / \phi_{0}$ and $\tau$ exhibit structure in their T-dependence near 40 K.  This structure is reminiscent of the T-dependence of the anomalous Hall coefficient $\sigma_{xy}$ that has been observed in thin films of SRO \cite{EHallEffect, monopoles, mathieu}.  For comparison, Fig. 4d reproduces  $\sigma_{xy}(T)$ reported in Ref. \cite{mathieu} The similarity between the T-dependence of AHE and parameters related to FMR suggests a correlation between the two types of response functions.  Recently Nagaosa and Ono \cite{lefthanded} have discussed the possibility of a close connection between collective spin dynamics at zero wavevector (FMR) and the off-diagonal conductivity (AHE).  At a basic level, both effects are nonzero only in the presence of both SO coupling and time-reversal breaking.   However, the possibility of a more quantitative connection is suggested by comparison of the Kubo formulas for the two corresponding functions.  The off-diagonal conductivity can be written in the form \cite{onodatopological},
\begin{equation}
\sigma_{xy} ( \omega ) = i \sum_{m,n,k} \frac{J^{x}_{mn} (k) J^{y}_{nm} ( k ) f_{mn} ( k )}{\epsilon_{mn}(k) \left[ \epsilon_{mn} (k) - \omega - i\gamma \right]},
\end{equation}
where $J^{i}_{mn} (k)$ is current matrix element between quasiparticle states with band indices $n,m$ and wavevector $k$.  The functions $\epsilon_{mn} (k)$ and $f_{mn} (k)$ are the energy and occupation difference, respectively, between such states, and $\gamma$ is a phenomenological quasiparticle damping rate.   FMR is related to the imaginary part of the uniform tranverse susceptibility, with the corresponding Kubo form,
\begin{equation}
\text{Im} \; \chi_{xy} ( \omega ) = \gamma \sum_{m,n,k} \frac{S^{x}_{mn} (k) S^{y}_{nm} (k) f_{mn} (k)} {\left[ \epsilon_{mn} (k) - \omega \right]^2 + \gamma^2},
\end{equation}
where $S^{i}_{mn}$ is the matrix element of the spin operator.  In general, $\sigma_{xy} ( \omega )$ and $\chi_{xy} ( \omega )$ are unrelated, as they involve current and spin matrix elements respectively.  However, it has been proposed that in several ferromagnets, including SRO, the \textit{k}-space sums in Eqs. 2 and 3 are dominated by a small number of band-crossings near the Fermi surface \cite{monopoles, souza}.  If the matrix elements $S^{i}_{mn}$ and $J^{i}_{mn}$ vary sufficiently smoothly with \textit{k}, then $\sigma_{xy} ( \omega)$ and $\chi_{xy} ( \omega )$ may be closely related, with both properties determined by the position of the chemical potential relative to the energy at which the bands cross.  Furthermore, as Gilbert damping is related to the zero-frequency limit of $\chi_{xy} ( \omega )$, i.e.,
\begin{equation}
\alpha = \frac{\Omega_{FMR}}{\chi_{xy} (0)} \frac{\partial}{\partial \omega} \lim_{\omega \to \infty} \text{Im} \; \chi_{xy} ( \omega ),
\end{equation}
and AHE is the zero-frequency limit of $\sigma_{xy} ( \omega)$, the band-crossing picture implies a strong correlation between $\alpha ( T )$ and $\sigma_{xy} ( T )$.

In conclusion, we have reported the observation of FMR in the metallic transition-metal oxide SrRuO$_{3}$. Both the frequency and damping coefficient are significantly larger than observed in transition metal ferromagnets.  Correlations between FMR dynamics and the AHE coefficient suggest that both may be linked to near Fermi surface band-crossings.  Further study of these correlations, as Sr is replaced by Ca, or with systematic variation in residual resistance, could be a fruitful approach to understanding the dynamics of magnetization in the presence of strong SO interaction.

\begin{acknowledgments}
This research is supported by the US Department of Energy, Office of Science, under contract No. DE-AC02-05CH1123.  Y.H.C. would also like to acknowledge the support of the National Science Council, R.O.C., under Contract No. NSC 97-3114-M-009-001.
\end{acknowledgments}


\begin{thebibliography}{25}
\expandafter\ifx\csname natexlab\endcsname\relax\def\natexlab#1{#1}\fi
\expandafter\ifx\csname bibnamefont\endcsname\relax
  \def\bibnamefont#1{#1}\fi
\expandafter\ifx\csname bibfnamefont\endcsname\relax
  \def\bibfnamefont#1{#1}\fi
\expandafter\ifx\csname citenamefont\endcsname\relax
  \def\citenamefont#1{#1}\fi
\expandafter\ifx\csname url\endcsname\relax
  \def\url#1{\texttt{#1}}\fi
\expandafter\ifx\csname urlprefix\endcsname\relax\def\urlprefix{URL }\fi
\providecommand{\bibinfo}[2]{#2}
\providecommand{\eprint}[2][]{\url{#2}}

\bibitem[{\citenamefont{\^Zuti\'c et~al.}(2004)\citenamefont{\^Zuti\'c, Fabian,
  and Das~Sarma}}]{spintronics}
\bibinfo{author}{\bibfnamefont{I.}~\bibnamefont{\^Zuti\'c}},
  \bibinfo{author}{\bibfnamefont{J.}~\bibnamefont{Fabian}}, \bibnamefont{and}
  \bibinfo{author}{\bibfnamefont{S.}~\bibnamefont{Das~Sarma}},
  \bibinfo{journal}{Rev. Mod. Phys.} \textbf{\bibinfo{volume}{76}},
  \bibinfo{pages}{323} (\bibinfo{year}{2004}).

\bibitem[{\citenamefont{Korenman and Prange}(1972)}]{korenman}
\bibinfo{author}{\bibfnamefont{V.}~\bibnamefont{Korenman}} \bibnamefont{and}
  \bibinfo{author}{\bibfnamefont{R.~E.} \bibnamefont{Prange}},
  \bibinfo{journal}{Phys. Rev. B} \textbf{\bibinfo{volume}{6}},
  \bibinfo{pages}{2769} (\bibinfo{year}{1972}).

\bibitem[{\citenamefont{Kambersk\'y}(1970)}]{kambersky}
\bibinfo{author}{\bibfnamefont{V.}~\bibnamefont{Kambersk\'y}},
  \bibinfo{journal}{Can. J. Phys.} \textbf{\bibinfo{volume}{48}},
  \bibinfo{pages}{2906} (\bibinfo{year}{1970}).

\bibitem[{\citenamefont{Slonczewski}(1996)}]{slonc}
\bibinfo{author}{\bibfnamefont{J.~C.} \bibnamefont{Slonczewski}},
  \bibinfo{journal}{J. Magn. Magn. Mater.} \textbf{\bibinfo{volume}{159}},
  \bibinfo{pages}{L1} (\bibinfo{year}{1996}).

\bibitem[{\citenamefont{Berger}(1996)}]{berger}
\bibinfo{author}{\bibfnamefont{L.}~\bibnamefont{Berger}},
  \bibinfo{journal}{Phys. Rev. B} \textbf{\bibinfo{volume}{54}},
  \bibinfo{pages}{9353} (\bibinfo{year}{1996}).

\bibitem[{\citenamefont{Luttinger and Karplus}(1954)}]{karplus}
\bibinfo{author}{\bibfnamefont{J.~M.} \bibnamefont{Luttinger}}
  \bibnamefont{and} \bibinfo{author}{\bibfnamefont{R.}~\bibnamefont{Karplus}},
  \bibinfo{journal}{Phys. Rev.} \textbf{\bibinfo{volume}{94}},
  \bibinfo{pages}{782} (\bibinfo{year}{1954}).

\bibitem[{\citenamefont{Brooks}(1940)}]{MCA}
\bibinfo{author}{\bibfnamefont{H.}~\bibnamefont{Brooks}},
  \bibinfo{journal}{Phys. Rev.} \textbf{\bibinfo{volume}{58}},
  \bibinfo{pages}{909} (\bibinfo{year}{1940}).

\bibitem[{\citenamefont{Klein et~al.}(1996)\citenamefont{Klein, Dodge, Ahn,
  Reiner, Mieville, Geballe, Beasley, and Kapitulnik}}]{anisotropydir}
\bibinfo{author}{\bibfnamefont{L.}~\bibnamefont{Klein}},
  \bibinfo{author}{\bibfnamefont{J.~S.} \bibnamefont{Dodge}},
  \bibinfo{author}{\bibfnamefont{C.~H.} \bibnamefont{Ahn}},
  \bibinfo{author}{\bibfnamefont{J.~W.} \bibnamefont{Reiner}},
  \bibinfo{author}{\bibfnamefont{L.}~\bibnamefont{Mieville}},
  \bibinfo{author}{\bibfnamefont{T.~H.} \bibnamefont{Geballe}},
  \bibinfo{author}{\bibfnamefont{M.~R.} \bibnamefont{Beasley}},
  \bibnamefont{and}
  \bibinfo{author}{\bibfnamefont{A.}~\bibnamefont{Kapitulnik}},
  \bibinfo{journal}{J. Phys. Cond.-Matt.} \textbf{\bibinfo{volume}{8}},
  \bibinfo{pages}{10111} (\bibinfo{year}{1996}).

\bibitem[{\citenamefont{Kostic et~al.}(1998)\citenamefont{Kostic, Okada,
  Collins, Schlesinger, Reiner, Klein, Kapitulnik, Geballe, and
  Beasley}}]{irconductivity}
\bibinfo{author}{\bibfnamefont{P.}~\bibnamefont{Kostic}},
  \bibinfo{author}{\bibfnamefont{Y.}~\bibnamefont{Okada}},
  \bibinfo{author}{\bibfnamefont{N.~C.} \bibnamefont{Collins}},
  \bibinfo{author}{\bibfnamefont{Z.}~\bibnamefont{Schlesinger}},
  \bibinfo{author}{\bibfnamefont{J.~W.} \bibnamefont{Reiner}},
  \bibinfo{author}{\bibfnamefont{L.}~\bibnamefont{Klein}},
  \bibinfo{author}{\bibfnamefont{A.}~\bibnamefont{Kapitulnik}},
  \bibinfo{author}{\bibfnamefont{T.~H.} \bibnamefont{Geballe}},
  \bibnamefont{and} \bibinfo{author}{\bibfnamefont{M.~R.}
  \bibnamefont{Beasley}}, \bibinfo{journal}{Phys. Rev. Lett.}
  \textbf{\bibinfo{volume}{81}}, \bibinfo{pages}{2498} (\bibinfo{year}{1998}).

\bibitem[{\citenamefont{Marshall et~al.}(1999)\citenamefont{Marshall, Klein,
  Dodge, Ahn, Reiner, Mieville, Antagonazza, Kapitulnik, Geballe, and
  Beasley}}]{marshall}
\bibinfo{author}{\bibfnamefont{A.~F.} \bibnamefont{Marshall}},
  \bibinfo{author}{\bibfnamefont{L.}~\bibnamefont{Klein}},
  \bibinfo{author}{\bibfnamefont{J.~S.} \bibnamefont{Dodge}},
  \bibinfo{author}{\bibfnamefont{C.~H.} \bibnamefont{Ahn}},
  \bibinfo{author}{\bibfnamefont{J.~W.} \bibnamefont{Reiner}},
  \bibinfo{author}{\bibfnamefont{L.}~\bibnamefont{Mieville}},
  \bibinfo{author}{\bibfnamefont{L.}~\bibnamefont{Antagonazza}},
  \bibinfo{author}{\bibfnamefont{A.}~\bibnamefont{Kapitulnik}},
  \bibinfo{author}{\bibfnamefont{T.~H.} \bibnamefont{Geballe}},
  \bibnamefont{and} \bibinfo{author}{\bibfnamefont{M.~R.}
  \bibnamefont{Beasley}}, \bibinfo{journal}{J. Appl. Phys.}
  \textbf{\bibinfo{volume}{85}}, \bibinfo{pages}{4131} (\bibinfo{year}{1999}).

\bibitem[{\citenamefont{Heinrich and Cochran}(1993)}]{heinrich}
\bibinfo{author}{\bibfnamefont{B.}~\bibnamefont{Heinrich}} \bibnamefont{and}
  \bibinfo{author}{\bibfnamefont{J.~F.} \bibnamefont{Cochran}},
  \bibinfo{journal}{Adv. Phys.} \textbf{\bibinfo{volume}{42}},
  \bibinfo{pages}{523} (\bibinfo{year}{1993}).

\bibitem[{\citenamefont{Hiebert et~al.}(1997)\citenamefont{Hiebert,
  Stankiewicz, and Freeman}}]{precession1}
\bibinfo{author}{\bibfnamefont{W.~K.} \bibnamefont{Hiebert}},
  \bibinfo{author}{\bibfnamefont{A.}~\bibnamefont{Stankiewicz}},
  \bibnamefont{and} \bibinfo{author}{\bibfnamefont{M.~R.}
  \bibnamefont{Freeman}}, \bibinfo{journal}{Phys. Rev. Lett.}
  \textbf{\bibinfo{volume}{79}}, \bibinfo{pages}{1134} (\bibinfo{year}{1997}).

\bibitem[{\citenamefont{Acremann et~al.}(2000)\citenamefont{Acremann, Back,
  Buess, Portmann, Vaterlaus, Pescia, and Melchior}}]{precession2}
\bibinfo{author}{\bibfnamefont{Y.}~\bibnamefont{Acremann}},
  \bibinfo{author}{\bibfnamefont{C.~H.} \bibnamefont{Back}},
  \bibinfo{author}{\bibfnamefont{M.}~\bibnamefont{Buess}},
  \bibinfo{author}{\bibfnamefont{O.}~\bibnamefont{Portmann}},
  \bibinfo{author}{\bibfnamefont{A.}~\bibnamefont{Vaterlaus}},
  \bibinfo{author}{\bibfnamefont{D.}~\bibnamefont{Pescia}}, \bibnamefont{and}
  \bibinfo{author}{\bibfnamefont{H.}~\bibnamefont{Melchior}},
  \bibinfo{journal}{Science} \textbf{\bibinfo{volume}{290}},
  \bibinfo{pages}{492} (\bibinfo{year}{2000}).

\bibitem[{\citenamefont{van Kampen et~al.}(2002)\citenamefont{van Kampen,
  Jozsa, Kohlhepp, LeClair, Lagae, de~Jonge, and Koopmans}}]{precession0}
\bibinfo{author}{\bibfnamefont{M.}~\bibnamefont{van Kampen}},
  \bibinfo{author}{\bibfnamefont{C.}~\bibnamefont{Jozsa}},
  \bibinfo{author}{\bibfnamefont{J.~T.} \bibnamefont{Kohlhepp}},
  \bibinfo{author}{\bibfnamefont{P.}~\bibnamefont{LeClair}},
  \bibinfo{author}{\bibfnamefont{L.}~\bibnamefont{Lagae}},
  \bibinfo{author}{\bibfnamefont{W.~J.~M.} \bibnamefont{de~Jonge}},
  \bibnamefont{and} \bibinfo{author}{\bibfnamefont{B.}~\bibnamefont{Koopmans}},
  \bibinfo{journal}{Phys. Rev. Lett.} \textbf{\bibinfo{volume}{88}},
  \bibinfo{pages}{227201} (\bibinfo{year}{2002}).

\bibitem[{\citenamefont{Shinagawa}(2000)}]{magnetooptics}
\bibinfo{author}{\bibfnamefont{K.}~\bibnamefont{Shinagawa}}, in
  \emph{\bibinfo{booktitle}{Magneto-optics}}, edited by
  \bibinfo{editor}{\bibfnamefont{S.}~\bibnamefont{Sugano}} \bibnamefont{and}
  \bibinfo{editor}{\bibfnamefont{N.}~\bibnamefont{Kojima}}
  (\bibinfo{publisher}{Springer-Verlag}, \bibinfo{address}{Berlin, Germany},
  \bibinfo{year}{2000}).

\bibitem[{\citenamefont{Ogasawara et~al.}(2005)\citenamefont{Ogasawara,
  Ohgushi, Tomioka, Takahashi, Okamoto, Kawasaki, and Tokura}}]{ogasawara}
\bibinfo{author}{\bibfnamefont{T.}~\bibnamefont{Ogasawara}},
  \bibinfo{author}{\bibfnamefont{K.}~\bibnamefont{Ohgushi}},
  \bibinfo{author}{\bibfnamefont{Y.}~\bibnamefont{Tomioka}},
  \bibinfo{author}{\bibfnamefont{K.~S.} \bibnamefont{Takahashi}},
  \bibinfo{author}{\bibfnamefont{H.}~\bibnamefont{Okamoto}},
  \bibinfo{author}{\bibfnamefont{M.}~\bibnamefont{Kawasaki}}, \bibnamefont{and}
  \bibinfo{author}{\bibfnamefont{Y.}~\bibnamefont{Tokura}},
  \bibinfo{journal}{Phys. Rev. Lett.} \textbf{\bibinfo{volume}{94}},
  \bibinfo{pages}{087202} (\bibinfo{year}{2005}).

\bibitem[{\citenamefont{Kise et~al.}(2000)\citenamefont{Kise, Ogasawara,
  Ashida, Tomioka, Tokura, and Kuwata-Gonokami}}]{slowingdown}
\bibinfo{author}{\bibfnamefont{T.}~\bibnamefont{Kise}},
  \bibinfo{author}{\bibfnamefont{T.}~\bibnamefont{Ogasawara}},
  \bibinfo{author}{\bibfnamefont{M.}~\bibnamefont{Ashida}},
  \bibinfo{author}{\bibfnamefont{Y.}~\bibnamefont{Tomioka}},
  \bibinfo{author}{\bibfnamefont{Y.}~\bibnamefont{Tokura}}, \bibnamefont{and}
  \bibinfo{author}{\bibfnamefont{M.}~\bibnamefont{Kuwata-Gonokami}},
  \bibinfo{journal}{Phys. Rev. Lett.} \textbf{\bibinfo{volume}{85}},
  \bibinfo{pages}{1986} (\bibinfo{year}{2000}).

\bibitem[{\citenamefont{Brown}(1963)}]{LLG}
\bibinfo{author}{\bibfnamefont{W.~F.} \bibnamefont{Brown}},
  \emph{\bibinfo{title}{Micromagnetics}} (\bibinfo{publisher}{Krieger},
  \bibinfo{year}{1963}).

\bibitem[{\citenamefont{Koopmans et~al.}(2000)\citenamefont{Koopmans, van
  Kampen, Kohlhepp, and de~Jonge}}]{koopmans}
\bibinfo{author}{\bibfnamefont{B.}~\bibnamefont{Koopmans}},
  \bibinfo{author}{\bibfnamefont{M.}~\bibnamefont{van Kampen}},
  \bibinfo{author}{\bibfnamefont{J.~T.} \bibnamefont{Kohlhepp}},
  \bibnamefont{and} \bibinfo{author}{\bibfnamefont{W.~J.~M.}
  \bibnamefont{de~Jonge}}, \bibinfo{journal}{Phys. Rev. Lett.}
  \textbf{\bibinfo{volume}{85}}, \bibinfo{pages}{844} (\bibinfo{year}{2000}).

\bibitem[{\citenamefont{Mathieu et~al.}(2004)\citenamefont{Mathieu, Asamitsu,
  Yamada, Takahashi, Kawasaki, Fang, Nagaosa, and Tokura}}]{mathieu}
\bibinfo{author}{\bibfnamefont{R.}~\bibnamefont{Mathieu}},
  \bibinfo{author}{\bibfnamefont{A.}~\bibnamefont{Asamitsu}},
  \bibinfo{author}{\bibfnamefont{H.}~\bibnamefont{Yamada}},
  \bibinfo{author}{\bibfnamefont{K.~S.} \bibnamefont{Takahashi}},
  \bibinfo{author}{\bibfnamefont{M.}~\bibnamefont{Kawasaki}},
  \bibinfo{author}{\bibfnamefont{Z.}~\bibnamefont{Fang}},
  \bibinfo{author}{\bibfnamefont{N.}~\bibnamefont{Nagaosa}}, \bibnamefont{and}
  \bibinfo{author}{\bibfnamefont{Y.}~\bibnamefont{Tokura}},
  \bibinfo{journal}{Phys. Rev. Lett.} \textbf{\bibinfo{volume}{93}},
  \bibinfo{pages}{016602} (\bibinfo{year}{2004}).

\bibitem[{\citenamefont{Klein et~al.}(2000)\citenamefont{Klein, Reiner,
  Geballe, Beasley, and Kapitulnik}}]{EHallEffect}
\bibinfo{author}{\bibfnamefont{L.}~\bibnamefont{Klein}},
  \bibinfo{author}{\bibfnamefont{J.~R.} \bibnamefont{Reiner}},
  \bibinfo{author}{\bibfnamefont{T.~H.} \bibnamefont{Geballe}},
  \bibinfo{author}{\bibfnamefont{M.~R.} \bibnamefont{Beasley}},
  \bibnamefont{and}
  \bibinfo{author}{\bibfnamefont{A.}~\bibnamefont{Kapitulnik}},
  \bibinfo{journal}{Phys. Rev. B} \textbf{\bibinfo{volume}{61}},
  \bibinfo{pages}{R7842} (\bibinfo{year}{2000}).

\bibitem[{\citenamefont{Fang et~al.}(2003)\citenamefont{Fang, Nagaosa,
  Takahashi, Asamitsu, Mathieu, Ogasawara, Yamada, Kawasaki, Tokura, and
  Terakura}}]{monopoles}
\bibinfo{author}{\bibfnamefont{Z.}~\bibnamefont{Fang}},
  \bibinfo{author}{\bibfnamefont{N.}~\bibnamefont{Nagaosa}},
  \bibinfo{author}{\bibfnamefont{K.}~\bibnamefont{Takahashi}},
  \bibinfo{author}{\bibfnamefont{A.}~\bibnamefont{Asamitsu}},
  \bibinfo{author}{\bibfnamefont{R.}~\bibnamefont{Mathieu}},
  \bibinfo{author}{\bibfnamefont{T.}~\bibnamefont{Ogasawara}},
  \bibinfo{author}{\bibfnamefont{H.}~\bibnamefont{Yamada}},
  \bibinfo{author}{\bibfnamefont{M.}~\bibnamefont{Kawasaki}},
  \bibinfo{author}{\bibfnamefont{Y.}~\bibnamefont{Tokura}}, \bibnamefont{and}
  \bibinfo{author}{\bibfnamefont{K.}~\bibnamefont{Terakura}},
  \bibinfo{journal}{Science} \textbf{\bibinfo{volume}{302}},
  \bibinfo{pages}{92} (\bibinfo{year}{2003}).

\bibitem[{\citenamefont{Onoda et~al.}(2008)\citenamefont{Onoda, Mishchenko, and
  Nagaosa}}]{lefthanded}
\bibinfo{author}{\bibfnamefont{M.}~\bibnamefont{Onoda}},
  \bibinfo{author}{\bibfnamefont{A.~S.} \bibnamefont{Mishchenko}},
  \bibnamefont{and} \bibinfo{author}{\bibfnamefont{N.}~\bibnamefont{Nagaosa}},
  \bibinfo{journal}{J. Phys. Soc. Jap.} \textbf{\bibinfo{volume}{77}},
  \bibinfo{pages}{013702} (\bibinfo{year}{2008}).

\bibitem[{\citenamefont{Onoda and Nagaosa}(2002)}]{onodatopological}
\bibinfo{author}{\bibfnamefont{M.}~\bibnamefont{Onoda}} \bibnamefont{and}
  \bibinfo{author}{\bibfnamefont{N.}~\bibnamefont{Nagaosa}},
  \bibinfo{journal}{J. Phys. Soc. Jap.} \textbf{\bibinfo{volume}{71}},
  \bibinfo{pages}{19} (\bibinfo{year}{2002}).

\bibitem[{\citenamefont{Wang et~al.}(2006)\citenamefont{Wang, Yates, Souza, and
  Vanderbilt}}]{souza}
\bibinfo{author}{\bibfnamefont{X.}~\bibnamefont{Wang}},
  \bibinfo{author}{\bibfnamefont{J.~R.} \bibnamefont{Yates}},
  \bibinfo{author}{\bibfnamefont{I.}~\bibnamefont{Souza}}, \bibnamefont{and}
  \bibinfo{author}{\bibfnamefont{D.}~\bibnamefont{Vanderbilt}},
  \bibinfo{journal}{Phys. Rev. B.} \textbf{\bibinfo{volume}{74}},
  \bibinfo{pages}{195118} (\bibinfo{year}{2006}).

\end{thebibliography}
\end{document}